# Malware Detection and Prevention using Artificial Intelligence Techniques

**Md Jobair Hossain Faruk*, Hossain Shahriar†, Maria Valero†, Farhat Lamia Barsha‡, Shahriar Sobhan¶ Md Abdullah Khan§, Michael Whitman¶, Alfredo Cuzzocrea‖, Dan Lo§, Akond Rahman‡ and Fan Wu****

*Department of Software Engineering and Game Development, Kennesaw State University, USA
†Department of Information Technology, Kennesaw State University, USA
‡Department of Computer Science, Tennessee Tech university, USA
§Department Computer Science, Kennesaw State University, USA
¶Institute for Cyber Workforce Development, Kennesaw State University, USA
‖iDEA Lab, University of Calabria, Rende, Italy and LORIA, Nancy, France
** Department of Computer Science, Tuskegee University, USA

*Abstract*—**With the rapid technological advancement, security has become a major issue due to the increase in malware activity that poses a serious threat to the security and safety of both computer systems and stakeholders. To maintain stakeholder's, particularly, end user's security, protecting the data from fraudulent efforts is one of the most pressing concerns. A set of malicious programming code, scripts, active content, or intrusive software that is designed to destroy intended computer systems and programs or mobile and web applications is referred to as malware. According to a study, naive users are unable to distinguish between malicious and benign applications. Thus, computer systems and mobile applications should be designed to detect malicious activities towards protecting the stakeholders. A number of algorithms are available to detect malware activities by utilizing novel concepts including Artificial Intelligence, Machine Learning, and Deep Learning. In this study, we emphasize Artificial Intelligence (AI) based techniques for detecting and preventing malware activity. We present a detailed review of current malware detection technologies, their shortcomings, and ways to improve efficiency. Our study shows that adopting futuristic approaches for the development of malware detection applications shall provide significant advantages. The comprehension of this synthesis shall help researchers for further research on malware detection and prevention using AI.**

*Index Terms*—**Artificial Intelligence, Malware, Detection System, Malware Prevention Technology, Software Security**

## I. Introduction

The computer has indeed come a long way since the scientific revolution from the 1500s until today with the contribution of a large number of greatest scientific scholars who conceptualized the concept of computing include John Napier, Blaise Pascal, Gottfried von Leibniz, Joseph Jacquard, Charles Babbage, Herman Hollerith, John V. Atanasoff, Clifford Berry, Konrad Zuse, Howard Aiken, John Mauchly, Presper Eckert, Remington Rand, Alan Turing, and John von Neumann [1]. As a result, computing technology emerges as the key part of almost every conceivable sector of human endeavor including technology, engineering, economy, education, and in general, every aspect of life [2], [3]. Particularly, since the beginning of the Information Age in the early 1980s, the world has witnessed a revolutionary period in world history that marked a rapid, societal shift from an industrialized, machine-based economy to one based on Information Technology [4]. Rapid scientific and technological advancement imposes numerous challenges, particularly around the domain of computer technology that has begun since the 1980s, and the number of different viruses rises over 40,000 so far and increased dramatically [5].

The first computer-based virus was discovered in 1982 on Apple II machines called "Elk Cloner" develop by a 15-year-old high school student Rich Skrenta [6]. A few years later in 1986, two brothers Basit Farooq Alvi and Amjad Farooq Alvi who wanted to prove that PC is not immune, write a pc based stealth virus called "Brain" [7]. The viruses were capable of replicating using floppy disks, inserting the infected floppy leads the PC to be infected, especially it's drive-by adopting three phases concepts, (i) Boot Loading (ii) Replication and (iii) Manifestation.

Since then, practices of using the malware-based application have been increasing rapidly by taking the advantage of the vulnerability of the software technology. In the early stage, computer viruses including Elk Cloner and Brain were not designed to damage or harm any computer system rather point on problems. However, malware changes the direction towards more and more destructive with the goal to disrupt computer operation, gather sensitive information, or gain access to private computer systems [8].

A large number of Malware has been discovered in the past few decades including The Morris Worm, ILOVEYOU, Melissa, Code Red, Sasser, Nimda, Slammer, Welchia, Commwarrior-A, Stuxnet, and CryptoLocker, and the creation of these viruses also mutated based on technological development [9]. All these computer viruses can affect any government, data center, laboratory, commercial, enterprise, organizational software application and propagate via normal use or download, installation of commercial software, mali-cious intent, or even by clicking a predefined link. According to a researcher [10], viruses must be introduced to a target computer system by persuading or tricking someone with legitimate access to install them on the system because computer viruses do not appear spontaneously. Once it appears, the result can be very devastating and a number of catastrophic loss was recorded ever since.

In order to prevent those attacks and catastrophes, scientists around the world attempt to design security tools and antivirus packages that are mainly used to prevent, detect, avoid, and remove viruses, Trojans, worms, etc, whereas firewalls are used to monitor incoming and outgoing connections [11]. The exact origins of the first antivirus are disputed, however, the first documented removal of a computer virus by an actual antivirus program was developed by a German computer security expert Bernd Robert in 1987 who came up with a program to get rid of Vienna, a virus that infected *.com* files on DOS-based systems, according to a report by Hotspot shield [12]. A number of manuals or automated malware detection and prevention systems are available for various platforms such as mobile devices, servers, gateways, and workstations that provide updates of the detection process and the prevention process starts with being proactive.

Being the technological wonderland, adopting futuristic techniques for the development of robust security tools and antivirus is today's priority. Advancing such fields shall contribute by detecting and preventing malware and keeping users away from unwanted software. And various industries need focus to protect user's data from not only malware attacks but also other security vulnerabilities and data breaches; the financial, aircraft, and healthcare domain is at the forefront of such a target which needs focus in order to protect the privacy and security of users and patient's records [4], [13]. Blockchain as a result can be adopted to secure data or records of various industries, healthcare, and transportation for example [14], [15] while Artificial Intelligence (AI) can be used to advance the fields of malware detection and prevention with the possibility to develop efficient, robust, and scalable malware recognition modules. According to Dr. Giovanni Vigna [16], co-founder and CTO of Lastline, Artificial intelligence (AI) cannot automatically detect and resolve every potential malware or cyber threat incident, but when it combines the modeling of both bad and good behavior, it can be a successful and powerful weapon against even the most advanced malware.

This paper pursues to present an analyzed synopsis of malware detection and prevention methodologies from the perspective of Artificial Intelligence. We will provide a detailed overview of AI applications in current malware detection systems, their limitations, scope to improve, and at last, will propose ideas to overcome current limitations. The primary contributions of the paper are as follows:

- We study potential malware detection and prevention techniques and investigate the potential of Artificial Intelligence (AI).
- We provide a comprehensive review on malware detection and prevention approaches based on Artificial Intelligence (AI).
- We discuss the limitations of existing methods and provide future research directions.

The rest of the paper is structured as follows: Section II defines the term Artificial Intelligence (AI) and Malware followed by Research Method and Related Work. Section IV gives a detailed presentation on techniques used for malware detection and prevention using AI. We discuss the limitation on existing approaches and future research directions in Section V. Finally, Section VI concludes the paper.

## II. Artificial Intelligence and Malware

In this section, we define Artificial Intelligence and Malware. In order to give a proper scenario, we provide the classification of both AI and Malware introduced by different researchers.

### A. *Artificial Intelligence*

Artificial intelligence (AI) is a technological phenomenon that all industries wish to exploit to benefit from efficiency gains and cost reductions because of its capability of replacing humans by undertaking intelligent tasks that were once limited to the human mind [17]. Nones *et al.* [18] define AI as the rapidly growing development of computer systems that are able to perform tasks that only human intelligence could ever accomplish. However, from the aspects of scholars, AI can be used for intelligence augmentation (IA) instead of being a replacement for the human mind which gives it strategic importance with identifying as a potential key driver of the current technological revolution. Thus, AI can be widely used in developing projects based on intellectual processes including the capacity for augmentation, conception, consciousness, investigation, enthusiastic information, thinking, arranging, innovation, and problem-solving in different sectors including Big Data, Security, Business Analytics and many more domains [19], [20].

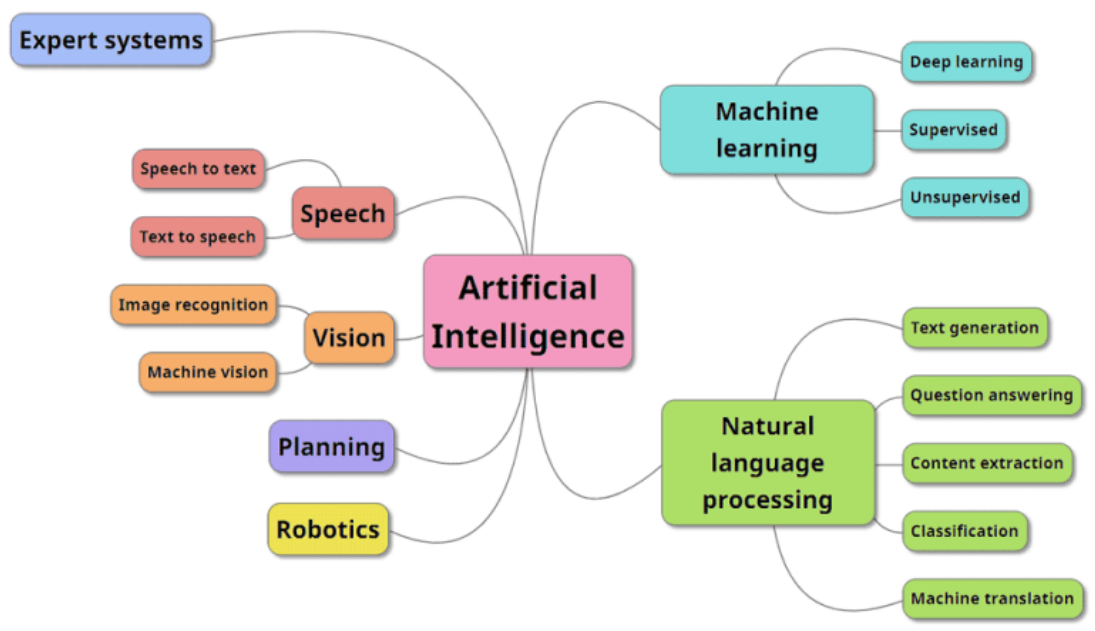

Fig. 1. Types and uses of Artificial Intelligence [21].

There are many types and ways AI can be achieved and machine learning is one of them which enables computers to imitate and adapt human-like behavior [22]. Fig. 1 illustrates the types and uses of Artificial Intelligence consists of Machine Learning and Natural Language Processing. Machine learning (MI) can be defined field of study that encompasses automatic computing procedures to computers to attain AI without being explicitly programmed based on logical or binary operations that learn a task from a series of examples [23]. Such understanding and learning through circumstances techniques of machine learning, AI can be the stepping stone for the application to prevent computer viruses and malware.

### B. Malware

Malware is a contraction of malicious programming codes, scripts, active content, or intrusive software that is designed to destroy intended computer systems and programs or mobile and web applications using different forms including computer viruses, worms, ransomware, rootkits, trojan, dialers, adware, spyware, keyloggers, or malicious Browser Helper Objects (BHOs) [24], [25]. Malware is the short form of malicious software or application which is not limited to computer system rather extend to the internet and related fields. Viruses and the Rise of the Internet grow gradually and significantly back, with only four hosts on the internet back in 1969, statistic shows the total number reached approximately 1.01 billion in 2019 [26], [27].

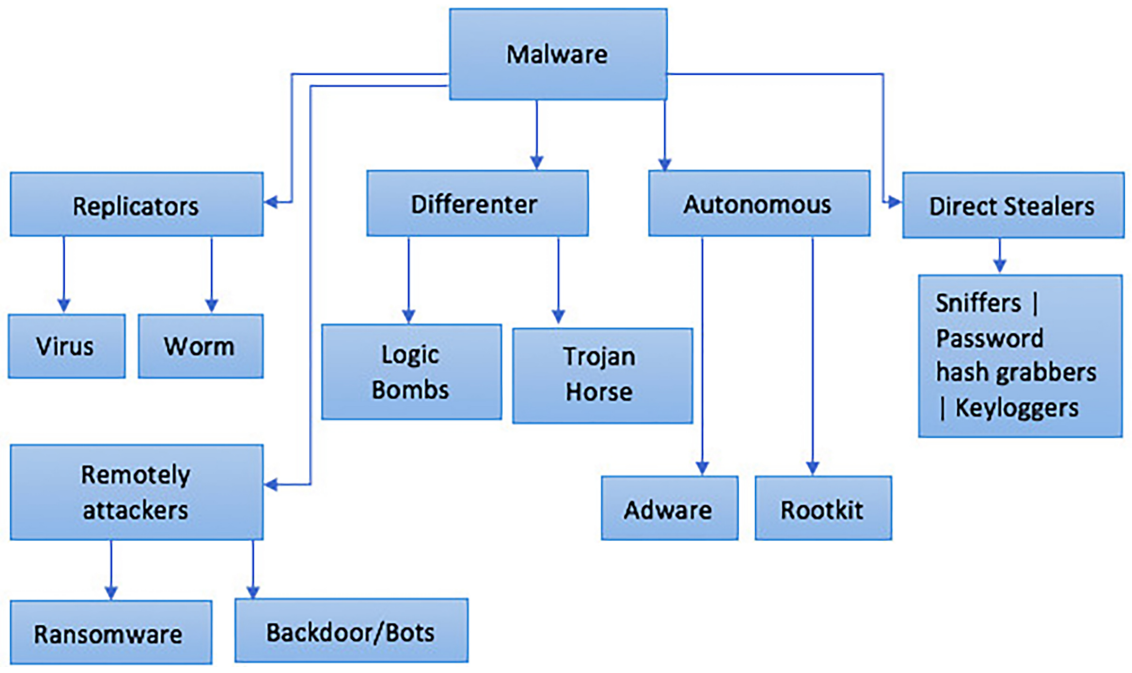

Fig. 2. Malware Classification by [28].

Any software purposefully designed for bad intention can be categorized as malware and it can be classified according to the purpose and method of propagation [29], [30]. Fig. 2 illustrates malware classification. A malware program can copy itself and infect a computer device without the permission or knowledge of the user, it can self-execute, if an infected file or a program is installed or shared with a new computer, the virus will automatically copy itself into the new computer and execute its code [30]. Such infected files or programs come from other sources, the internet in general, downloading files from malicious websites, or clicking on a malicious link in particular.

## III. Research Method and Related Work

This section will discuss the adopted research methodology and related work. We first define effective research methodology based on the research topic and broader domain followed by related work.

### A. Research Methodology

In order to carry out the study on Malware Detection Techniques using AI, we utilize a systematic literature review [31]. The main purpose of the systematic review is to identify, study, and investigate the suitable existing approaches. We first carried out a "Search Process" to identify potential research papers from the scientific databases using pre-selected search keywords or strings including "Artificial Intelligence" AND ("Malware" AND "Detection" OR "Prevention") OR "AI". We had to identify these search strings to avoid findings from non-related research papers and those keywords are based on the Malware and Artificial Intelligence related terms and their derivatives, acronyms, and widely used synonyms. Besides, among various scientific databases, we used three digital database sources including (i) IEEE Xplore (ii) ScienceDirect, and (iii) Springer Link. Choosing these three bibliographic databases, our goal is to identify research papers that were published in reputable conferences, journals, and books.

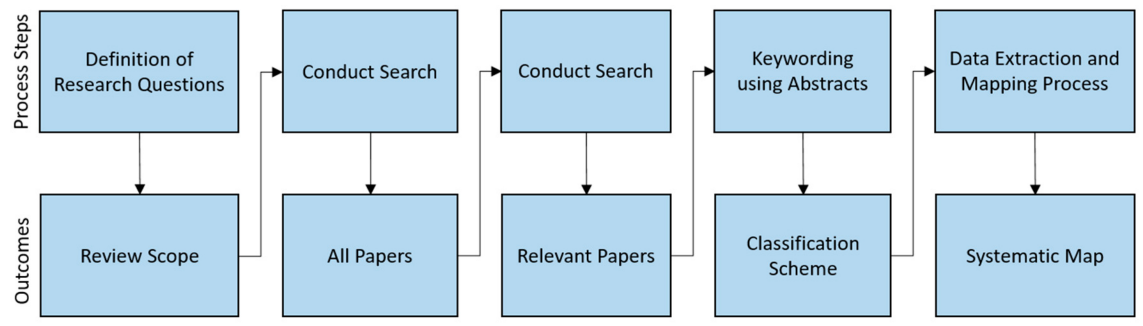

Fig. 3. Paper Classification Process [32]

TABLE I
Generalized Table for Search Criteria

| Database | Initial Search | Total Inclusion |
|---|---|---|
| IEEE Xplore | 42 | 8 |
| ScienceDirect | 111 | 5 |
| Springer Link | 35 | 3 |
| Total | 196 | 16 |

TABLE II
Overview of Exclusion and Inclusion

| Condition of Exclusion and Inclusion | | |
|---|---|---|
| Category | Condition (Inclusion) | Condition (Exclusion) |
| Type of Papers | Malware Detection, Prevention, Artificial Intelligence | Other studies than aforementioned topics |
| Duplicate Papers | Papers are not duplicated in different databases | Similar papers in different databases |
| Relativity | Papers and proposed approaches are similar aspects | Studies that do not depict expected aspects |
| Text Availability | Studies that are available in the full format | Studies are not available fully |

We adopt the paper classification process from Anushree Tandon *et al.* [32] depicted in Fig. 3 We filtered based on time restrictions that we set between 2016 to 2022 to search studies published. We also filtered publication topics including *Systems and Data Security* for Springer Link and *Publication Topics* for IEEE Explore and *Computer Science and Security* for ScienceDirect. A total of 196 studies were found during the initial search (IEEE Xplore 42, ScienceDirect 43, and Springer Link 111). Once the search processes are completed, we have gone through a screening process for finding relevant papers based on the paper title at first followed by reading and understanding the abstract and conclusion from screened papers. In order to apply the inclusion and exclusion criteria, we set a number of exclusion criteria including (i) duplicate papers (ii) full-text availability, and (iii) papers that are not related to Malware Detection and Prevention shown in table I - II.

### B. Related Work

- The purpose of Malware detection is to protect the system from various kinds of malicious attacks by following the policy of detection and prevention. There are various existing algorithms to detect malware, however, with the advancement of malware technology, the adoption of Artificial Intelligence is crucial for efficient, and robust malware prevention applications. In order to detection of malware, finding the malicious source code is the very first step.
- Researchers [33], proposed a technique called SourceFinder to identify malware source code repositories which is the largest malware source code database. The research showed that the proposed approach identifies malware repositories with 89 % precision and 86 % recall while it identifies 7504 malware source code repositories using SourceFinder followed by analyzing properties and characteristics of repositories.

  The use of Machine learning techniques to detect malware is a common practice. Likely many others. Niharika Sharma [34] presents a detailed analysis of the static, dynamic, and hybrid methods with an evaluation of malware detection techniques. The author also facilitates the detection process by combining machine learning and data mining techniques. Besides, the paper evaluates various data mining and machine learning-based malware detection approaches.
- Sanjay Sharma *et al.* [35] proposes an approach based on opcode occurrence to detect malware using machine learning techniques. The researchers also use a dataset from the Kaggle Microsoft malware classification challenge dataset and evaluate five classifiers including LMT, REPTree, Random Forest, NBT, J48Graft. A demonstration indicates that the proposed approach capable of detecting the malware with almost 100% accuracy.
- Despite challenges of applying machine learning in intrusion detection like unconventional computing paradigms and unconventional evasion techniques, Sherif Saad *et al.* [36], presents three critical problems that limit the success of malware detectors using machine learning techniques. The researchers also discuss the crucial behavioral analysis that shall dominate the next generation antimalware systems followed by proposing possible solutions to overcome the constraints.
- Apart from machine learning, other techniques are also used in malware detection like cloud computing, network-based detection system, virtual machine, or the use of hybrid methods and technologies. Nowadays Deep learning and Artificial are actively applied in malware detection. Irina Baptista *et al.* [37] introduces a novel approach for malware detection by utilizing binary visualization and self-organizing incremental neural networks. A demonstration was conducted on detecting malicious payloads in various file types including Portal Document File *.pdf* and Microsoft Document Files *.doc* files where the experimental results indicate a detection accuracy of 91.7% and 94.1% for ransomware respectively. According to the authors, the proposed technique performed well with an incremental detection rate, allowing for efficient real-time identification of unknown malware.
- In a separate study, Syam and Vankata [38] propose a detection way where a virtual analyst was developed by using Artificial Intelligence to defend threats and take appropriate measurements. The researchers categorize supervised and unsupervised data, and later converted unsupervised data to supervised data with the help of analyst feedback and then auto-update the algorithm. It evolves the algorithm by utilizing Active Learning Mechanism itself over time and becomes more efficient, strong.
- A group of researchers from Kennesaw State University [39], propose a novel Bayesian optimization-based framework for automated hyperparameter optimization, resulting in the optimum DNN architecture. The research evaluates the NSL-KDD, a benchmark dataset for network intrusion detection and the demonstration results indicate the framework's efficacy, resulting DNN architecture performs detects significantly higher intrusion in terms of accuracy, precision, recall, and f1-score. The BO-GP based approach outperforms the random search optimization-based method where BO-GP achieved the highest accuracy- 82.95%, and 54.99% for KDDTest+, and KDDTest-21 datasets, respectively.

## IV. Malware Detection Using AI

In this section, we discuss Artificial Intelligence-based techniques to detect malware, limitations of currently used strategies, and ways to overcome the shortcoming to improve performance.

### A. Malware Detection Techniques

Researchers develop malware detection systems, keep track of the malicious programs and benign software towards an-alyzing those applications in order. Malware detection techniques can be classified into three categories including (i) signature-based, (ii) anomaly-based, and (iii) heuristic-based. In this section, we discuss the malware detection systems and present the result and possible limitations.

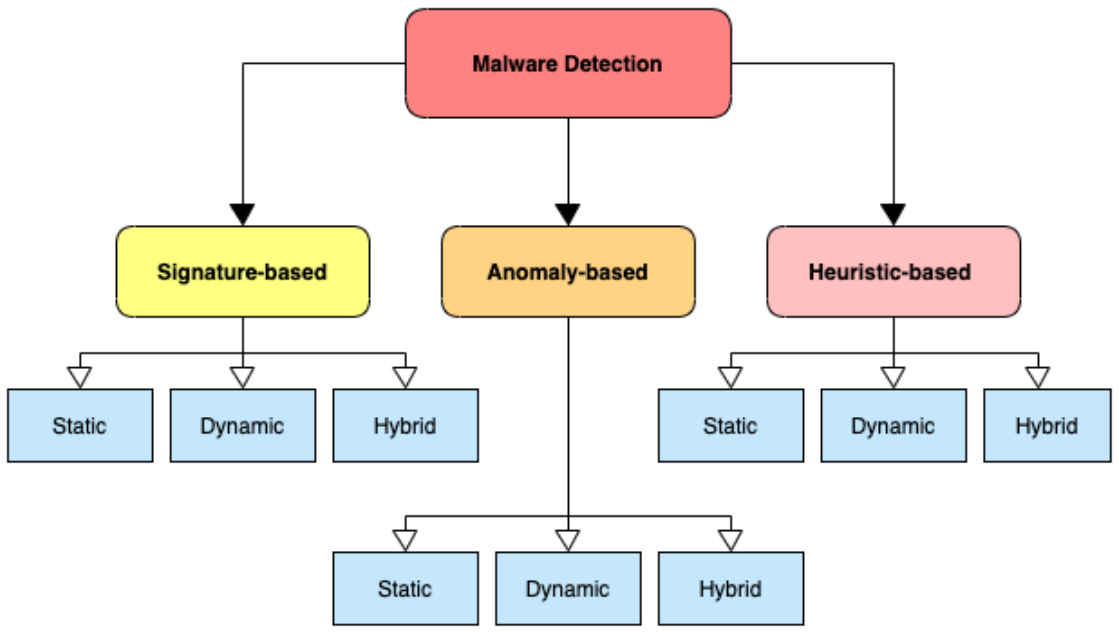

Fig. 4. Classification of Malware Detection Techniques

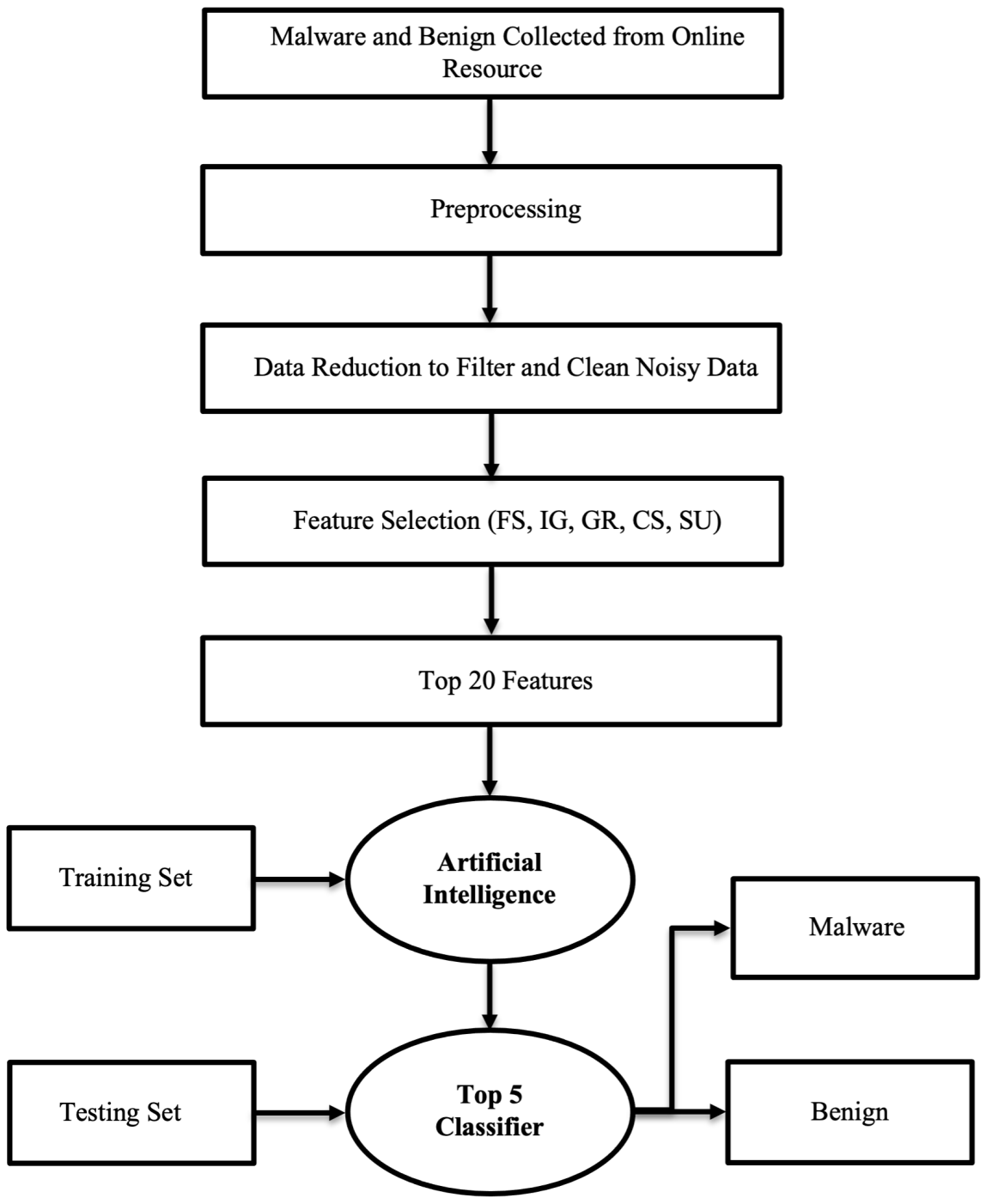

Fig. 5. Flow chart of AI Based Unknown Malware Detection Techniques

Fig. 4 illustrates the techniques for malware detection that work in a flow including data processing, feature selection, classifier training, and malware detection. The process begins with collecting datasets from the Kaggle website consisting of malware and benign web application. By adopting AI technology, the development of malware detection systems shall be in a way that will process malware datasets, and analyze malware to understand its feature. Fisher Score (FS), Chi-Square (CS), Information Gain (IG), Gain Ratio (GR), and Uncertainty Symmetric (US) are used to select 20 features. The system shall train the classifier by comparing different classifiers on FS, CS, IG, GR, and US to detect unknown malware.

Implementing different types of classifiers to develop malware detection and prevention systems shall provide better and using AI shall bring a significant advantage to detect and prevent unknown malicious activities [35]. In Fig. 5, we display a flowchart of unknown malware detection using artificial intelligence. In this section, we provide a detailed review on each method of Malware Detection.

- Signature-based Detection Technique: The signature-based detection method consists of four components as depicted in Fig. 6 is a term that helps in identifying and detecting attacks by looking for specific patterns [40]. In a signature-based method, developers use a database containing signatures of viruses, scan the file, and evaluate information with that database for detecting malware in the database. If the information matches with the database's data that means the file contains viruses. The primary advantage of this method is effective for the known malware, however, it has limitations in detecting unknown malware [41]. Fig. 7 shows Intrusion Detection System (IDS) keeps a statistical model of traffic that also can be referred to a database, IDS accepts traffic from various sources and matches it with statistical traffic to find out whether it is malicious or not and then provide the result to an administrator.

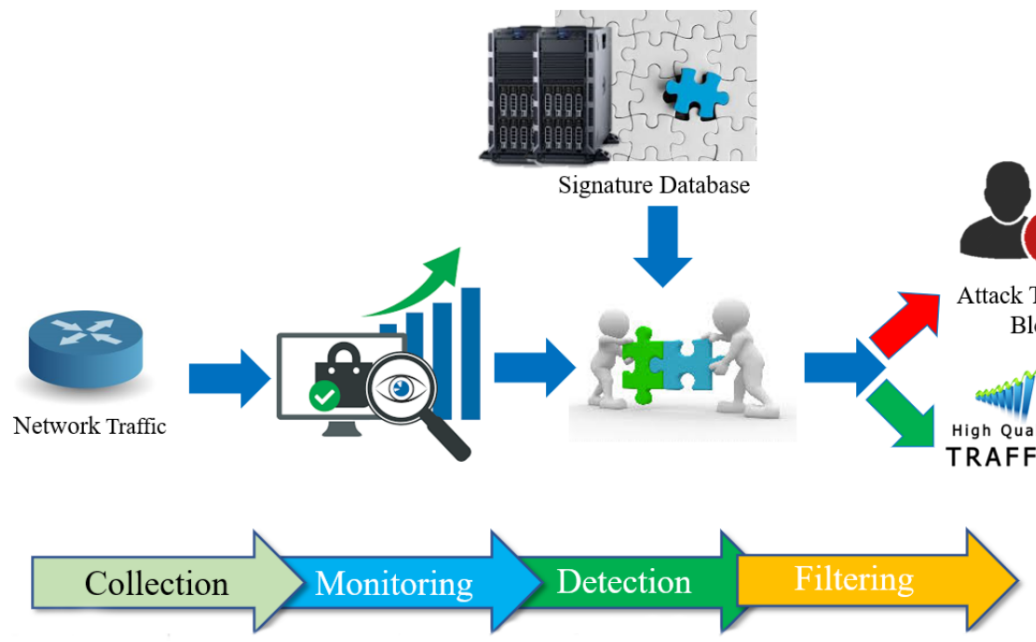

Fig. 6. Methodology used in Signature based IDS [42]

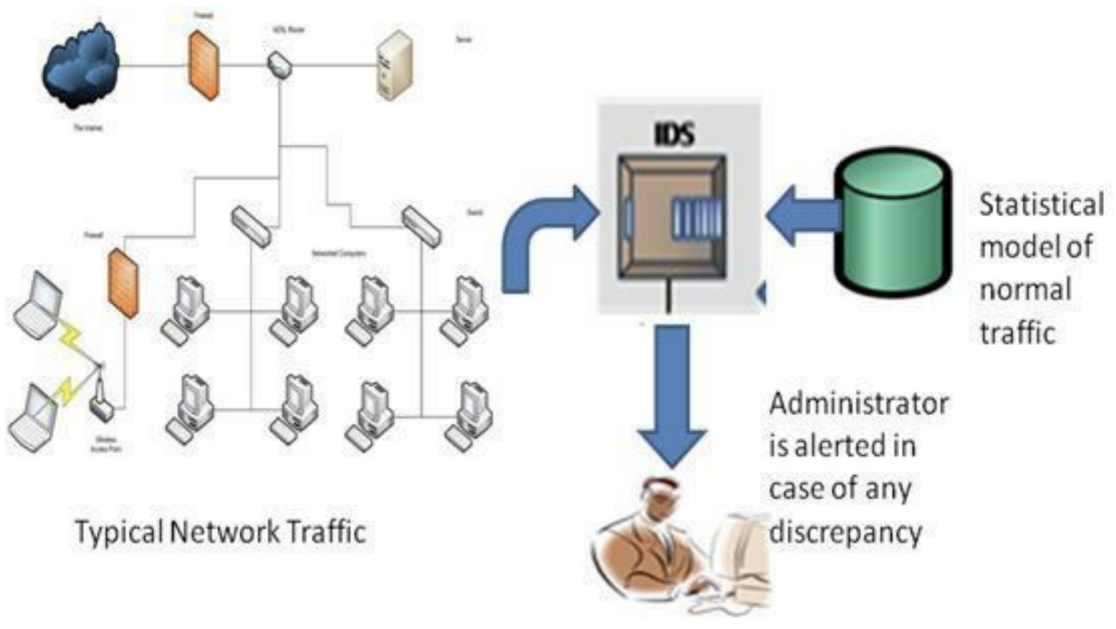

Fig. 7. Signature based Intrusion Detection System (IDS) [38]

- Anomaly-based Detection Technique: Anomaly-based network intrusion detection plays a vital role in addressing security issues and protecting networks against malicious activities [43]. Anomaly-based methods address the limitations of signature-based techniques by enabling to detect any of known or unknown malware by applying classification techniques over activities of a system for malware detection. Such transformation from pattern-based detection to a classification-based approach to identify normal or anomalous behavior gives an advantage of detecting malware activities [44].

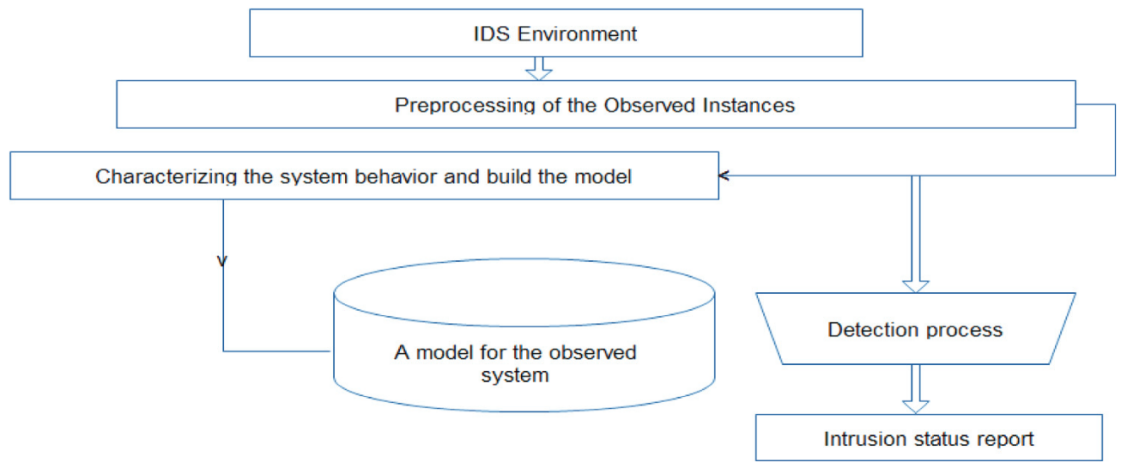

Fig. 8. Common anomaly-based network IDS [43]

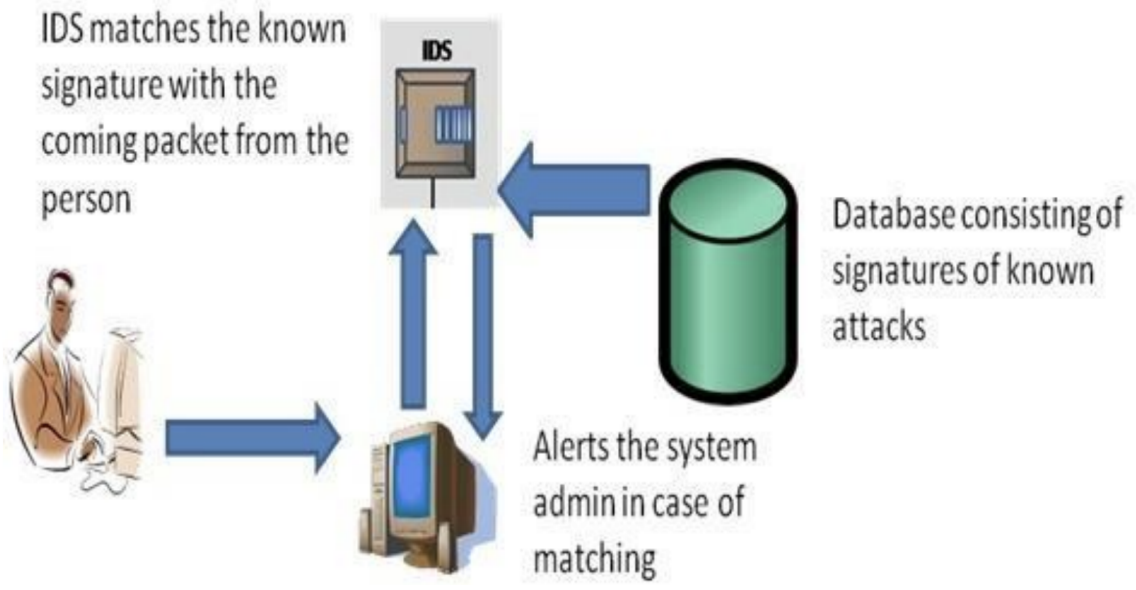

Fig. 9. Anomaly Based IDS [38]

Fig. 8 depicts the anomaly-based Network Intrusion Detection System (IDS) where the functional stages are normally adopted in the anomaly-based network intrusion detection systems (ANIDS). On the other hand, Fig. 9 illustrates a connection with a database consists of the signature of known attacks, with the common signatures coming from different packets with that database, an alert is sent to the system admin if the unknown signature matches with known signature mean malware detected.

- Heuristic-based Detection Technique: Applying Artificial Intelligence over the signature and anomaly-based detection systems improve the efficiency of malware detection. However, in order to adopt environmental change and improve prediction ability, a machine learning algorithm named genetic algorithm along with neural network was applied over malware detection system to improve the classification method. The algorithm applies characteristics such as inheritance, selection, and combination that give the advantage to attain optimum solutions from multiple directions without any previous knowledge about the system [45]. The combination of statistical and mathematical techniques improves the heuristic method from previous methods. Fig. 9 represents the features of Heuristic Methods.

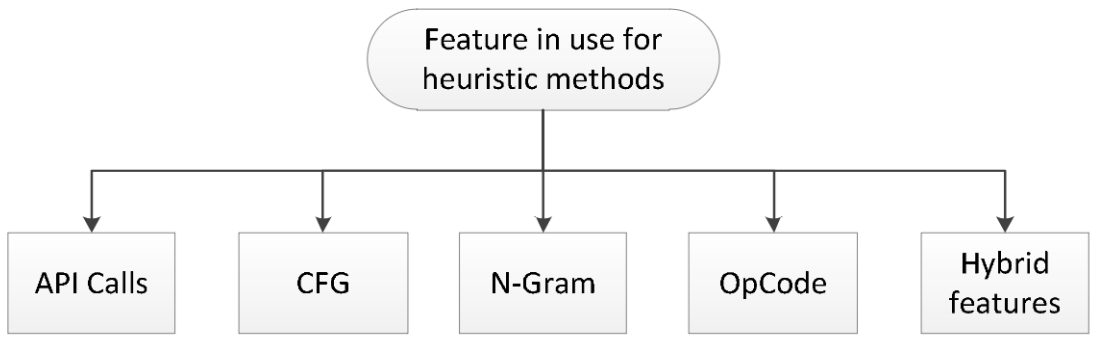

Fig. 10. Heuristic Methods Features [46]

## B. *Malware Detection by Adopting AI*

The continued evolution and diversity of malware constitutes a major threat in modern systems and existing security defenses are ineffective to mitigate the skills and imagination of cyber-criminals necessitating the development of novel solutions [47]. Besides, Artificial Intelligence (AI) is rapidly evolving and advances in AI enable remarkable results in many application areas and such advancement of the fields of AI shall be significant in the development of efficient anti-malware systems to overcome limitations of existing prevention technology. In this section, we discuss the malware detection technique sing AI and present the result and possible limitations.

Tal Garfinkel and Mendel Rosenblum [48] proposed a virtual machine monitoring approach to detect malicious software. An architectural framework (Fig. 11) was introduced that maintains the transparency of the host-based Intrusion detection system (IDS) but for larger attack resistance they kept the IDS away from the host. The evaluation indicates to achieve distinct ability to moderate interactions between host and principal software by using virtual machine monitor. However, the risk of errors and tamper resistance is the limitation of the proposed approach.

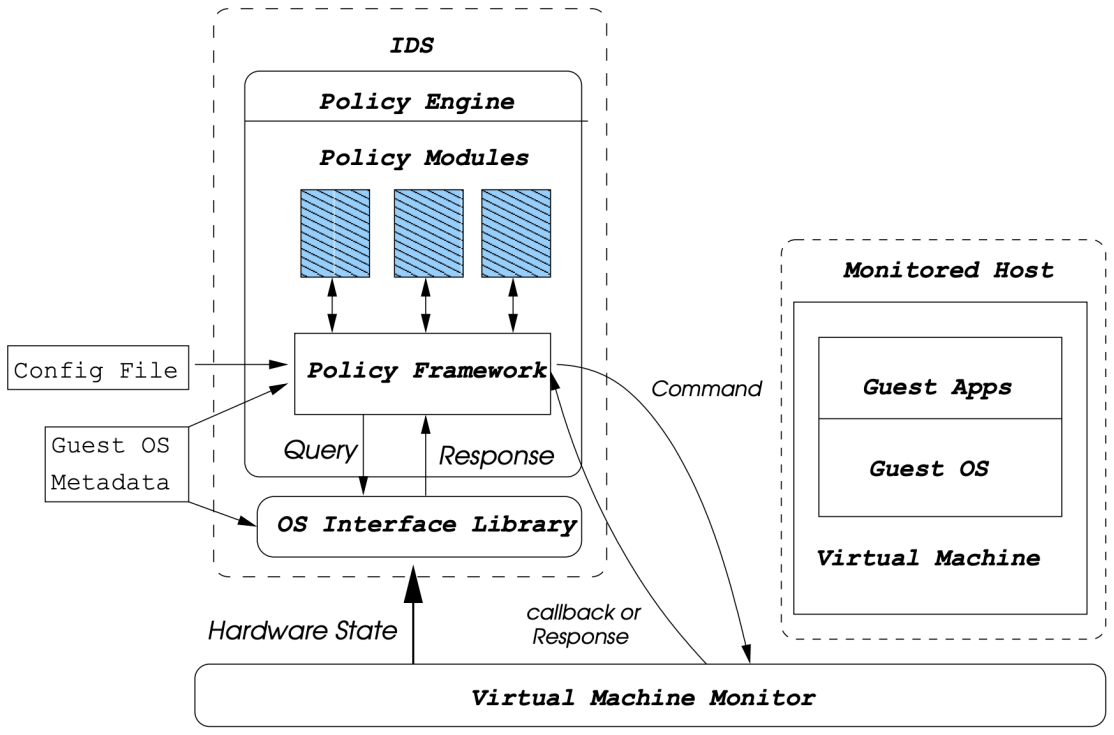

Fig. 11. A High-Level View of the VMI-Base [48]

Shanxi Li *et al.* [49] proposes a malware classifier based on graph convolutional network, designed to adapt to the dif-ference of malware characteristics. The approach first extracts the API call sequence from the malware code and generates a directed cycle graph followed by extracting the feature map of the graph, and designing a classifier based on graph convolutional network using the Markov chain and principal component analysis method. The method also analyzes and compares the performance of the method. Fig. 12 illustrates the GCN-based malware detection system framework. An evaluation result indicates the highest accuracy is 98.32. % which is superior compared to other existing methods in terms of FPR and accuracy

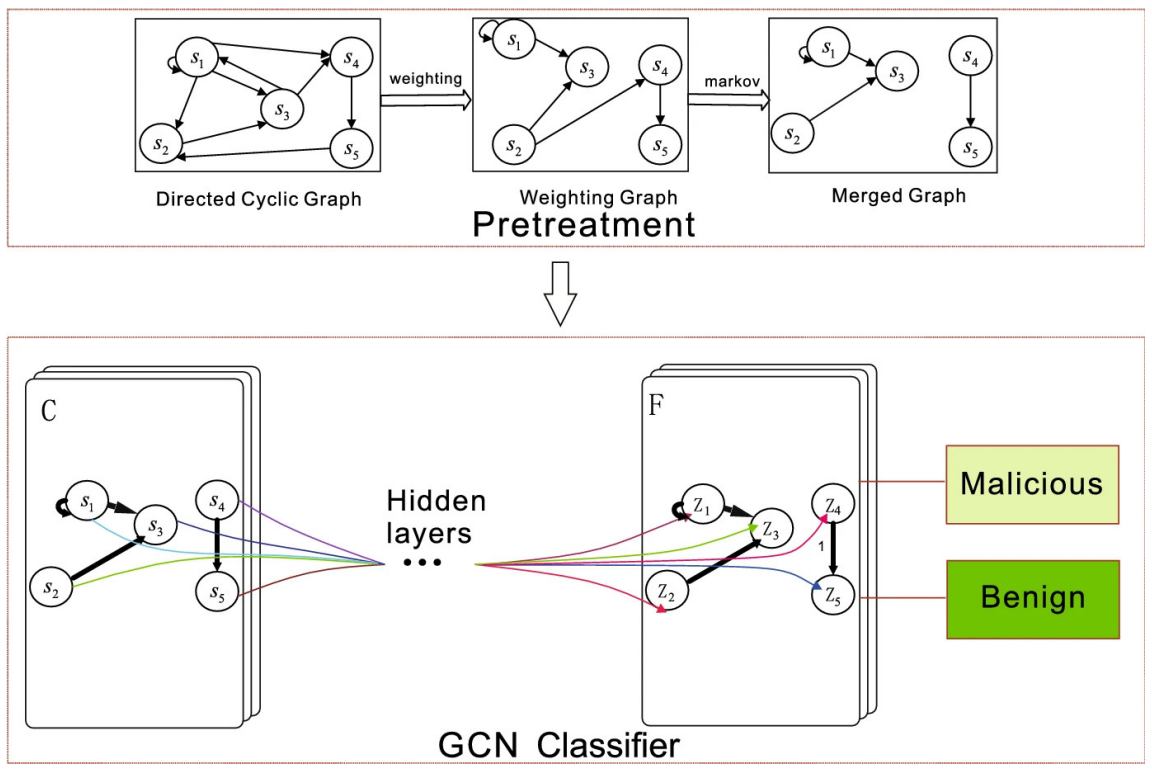

Fig. 12. GCN-based malware detection system framework [49]

Other than that, Long Wen and Haiyang Yu [50] propose a machine learning-based lightweight system with the goal of identifying unknown malware on Android devices. The proposed approach extract features based on static analysis and dynamic analysis. The researchers also introduce a new feature selection algorithm PCA-RELIEF towards disposing of the raw features. Fig. 13 depicts the architecture of machine learning-based Android malware detection. The demonstration achieved better performance with a higher detection rate and lower error detection.

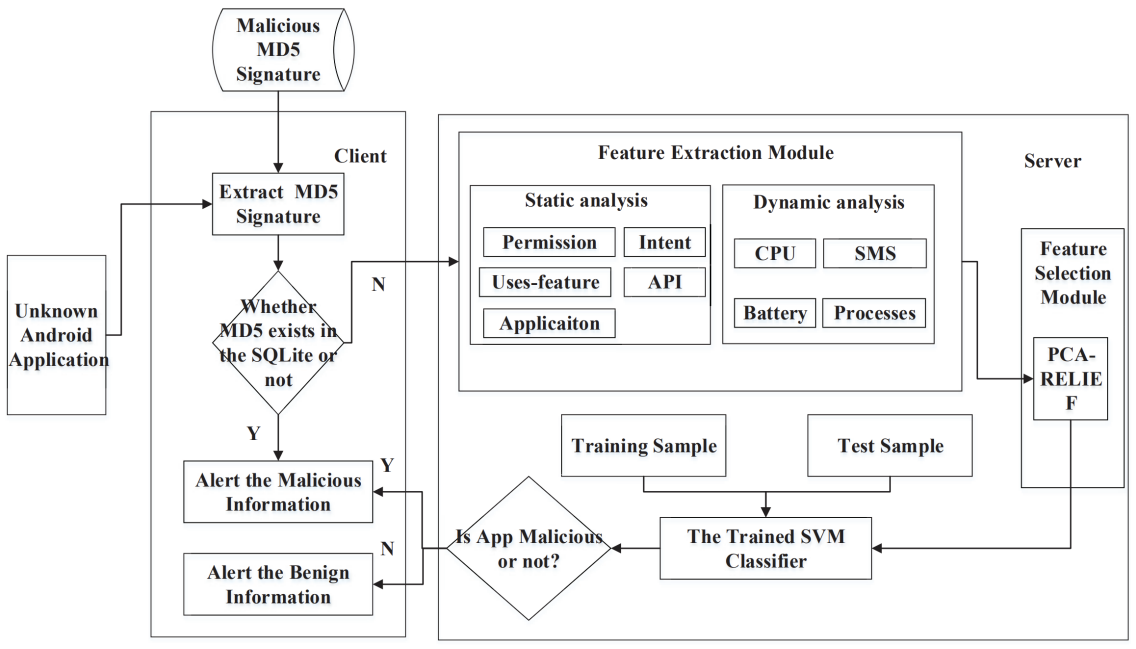

Fig. 13. The architecture of machine learning-based Android malware detection [50]

## V. Discussion and Limitations

Previously we discussed different kinds of malware detection techniques that contribute to solving the limitations of previous techniques. Analyzing the limitations of detection systems is vital to deal with novel techniques for malware detection and prevention. In this section, we aim to discuss the limitation of the presented approaches with suggestions to improve those limitations.

The primary limitation of the static signature-based method is the failure in detecting unknown malware activities. Update the database regularly may solve the issue temporarily because of some special viruses that have the capability of modifying the code after infecting any system. Such issues were addressed in the Generic signature scanning-based method that can detect unknown viruses; however, the method is unable to remove the affected files from the directory.

Heuristic analysis is divided into two parts- static and dynamic where it performs code mapping which is a difficult process as some characteristics of a virus can be implemented in various ways. Dynamic heuristic analysis is better comparably with static analysis regardless of its slow process. One of the limitations of dynamic processes is the failure not to detect certain active viruses in a certain situation. For instance, performing any operation by the user may interrupt the heuristic dynamic analysis. Integrity checking may solve the limitation of dynamic heuristic analysis, able to detect viruses with certainty if accuracy can be ignored because of its record in failure cases. Other than that, integrity checking always assumes that the starting state of a file is unaffected but this can be false often.

Malware detection techniques are working simultaneously to detect malicious software applications. In order to improve the efficiency of malware detection techniques, improvement of existing limitations is a major fact and dynamic solutions are needed to reduce malware feature analysis time, and more sophisticated approaches should be applied to detect malicious activities. Utilizing artificial intelligence (AI) technology in the development of both malware detection and prevention needs to be increased to deal with intelligent malware that has grown in recent years.

## VI. Conclusion

Malware or malicious applications may cause catastrophic damages to not only computer systems but also data centers, web, and mobile applications to various industries; particularly, financial and healthcare institutes. Ensuring the safety of stakeholders' data from malicious entities is a major challenge that leads us towards the concept of malware detection and prevention. Artificial Intelligence (AI) can be an effective solution that we can adopt for the development of Anti-Malware Systems. Having such direction, this study presented a detailed review of malware detection techniques and approaches. At first, we attempted to provide a clear overview of malware, artificial intelligence, and its narration. An overview of existing malware detection systems was discussed in section III (B) followed by identifying the limitations of existing applications. Likely every system, the malware detection approaches also consist of a number of limitations along with facilities and improvements from its previous version. So far, our findings indicate that AI can be utilized as a promising domain for the development of anti-malware systems for detecting and preventing malware attacks or security risks of software applications towards a technological wonderland. To draw a conclusion, we discuss scores of ideas to overcome the identified limitations and aim to continue our effort explicitly towards significant accomplishments around the domain of Malware Detection and Prevention.

## Acknowledgment

The work is partially supported by the U.S. National Science Foundation Awards #2100134, #2100115, #1723578, and #1723586. Any opinions, findings, and conclusions or recommendations expressed in this material are those of the authors and do not necessarily reflect the views of the National Science Foundation.